\documentclass[aps,pra,epsfigure,twocolumn,superscriptaddress]{revtex4-1}
\usepackage[colorlinks=true,linkcolor=blue,urlcolor=blue,citecolor=blue,pdfusetitle]{hyperref}
\usepackage[utf8]{inputenc}
\usepackage[english]{babel}
\usepackage{amsmath}
\usepackage[caption = false]{subfig}
\usepackage{graphicx,epstopdf}
\usepackage{blindtext}
\usepackage{lipsum}
\usepackage{amsfonts}
\usepackage{bbm}
\usepackage{amssymb}
\usepackage{enumerate}
\usepackage{color}
\usepackage{latexsym}
\usepackage{times,txfonts}

\newcommand{\Ccal}{\mathcal{C}}

\newcommand{\Pcal}{\mathcal{P}}

\newcommand{\Lcal}{\mathcal{L}}

\newcommand{\bra}[1]{\langle #1 |}
\newcommand{\ket}[1]{ |#1 \rangle}

\newcommand{\interpro}[2]{\langle #1 | #2 \rangle}

\newcommand{\trs}[1]{ \text{Tr} \{ #1 \}}

\begin{document}

%\title{Stable charging process and self-discharging in three-level quantum batteries}
\title{Stable adiabatic quantum batteries}

\author{Alan C. Santos}
\email{ac\_santos@id.uff.br}
\affiliation{Instituto de F\'{i}sica, Universidade Federal Fluminense, Av. Gal. Milton Tavares de Souza s/n, Gragoat\'{a}, 24210-346 Niter\'{o}i, Rio de Janeiro, Brazil}

\author{Bar\i\c{s} \c{C}akmak}
\email{baris.cakmak@eng.bau.edu.tr}
\affiliation{College of Engineering and Natural Sciences, Bah\c{c}e\c{s}ehir University, 34353 Be\c{s}ikta\c{s}, Istanbul, Turkey}

\author{Steve Campbell}
\email{steve.campbell@tcd.ie}
\affiliation{School of Physics, Trinity College Dublin, Dublin 2, Ireland}

\author{Nikolaj T. Zinner}
\email{zinner@phys.au.dk}
\affiliation{Department of Physics and Astronomy, Aarhus University, DK-8000 Aarhus C, Denmark}
\affiliation{Aarhus Institute of Advanced Studies, Aarhus University, DK-8000 Aarhus C, Denmark}

\begin{abstract}
With the advent of quantum technologies comes the requirement of building quantum components able to store energy to be used whenever necessary, i.e. quantum batteries. In this paper we exploit an adiabatic protocol to ensure a stable charged state of a three-level quantum battery which allows to avoid the spontaneous discharging regime. We study the effects of the most relevant sources of noise on the charging process and, as an experimental proposal, we discuss superconducting transmon qubits. In addition we study the self-discharging of our quantum battery where it is shown that spectrum engineering can be used to delay such phenomena.
\end{abstract}

\maketitle

\section{Introduction}
In recent years, building upon the advancements in quantum thermodynamics~\cite{JPA_Goold,CP_Vinjanampathy,arXiv_Alicki, Deffner2019book}, there is an increasing interest in developing new quantum devices with potential application to emerging quantum technologies such as quantum information processing \cite{xiang2013,georgescu2014,acin2018,krantz2019}, including components like quantum transistors~\cite{hwang2009,chen2013,Marchukov:16,Bacon:17,Ponte:18,sun2018} and quantum diodes~\cite{Landi:14,Balachandran:18,kargi2018,Pereira:19}. In this direction, developing strategies to store energy to be consumed by quantum devices has been a major issue to be addressed and therefore heralded the introduction of quantum batteries by R. Alicki and M. Fannes~\cite{Alicki:13}, and has subsequently developed into a significant field of research~\cite{PRL2013Huber, CampbellBatteries, Binder:15, PRL2017Binder, FriisQuantum, PRB2019Batteries, PRE2019Batteries, PRL2019Barra, Campaioli:18,Ferraro:18}.

\begin{figure}[b]
\centering
(a) \hskip0.5\columnwidth (b)\\
\includegraphics[width=0.5\columnwidth]{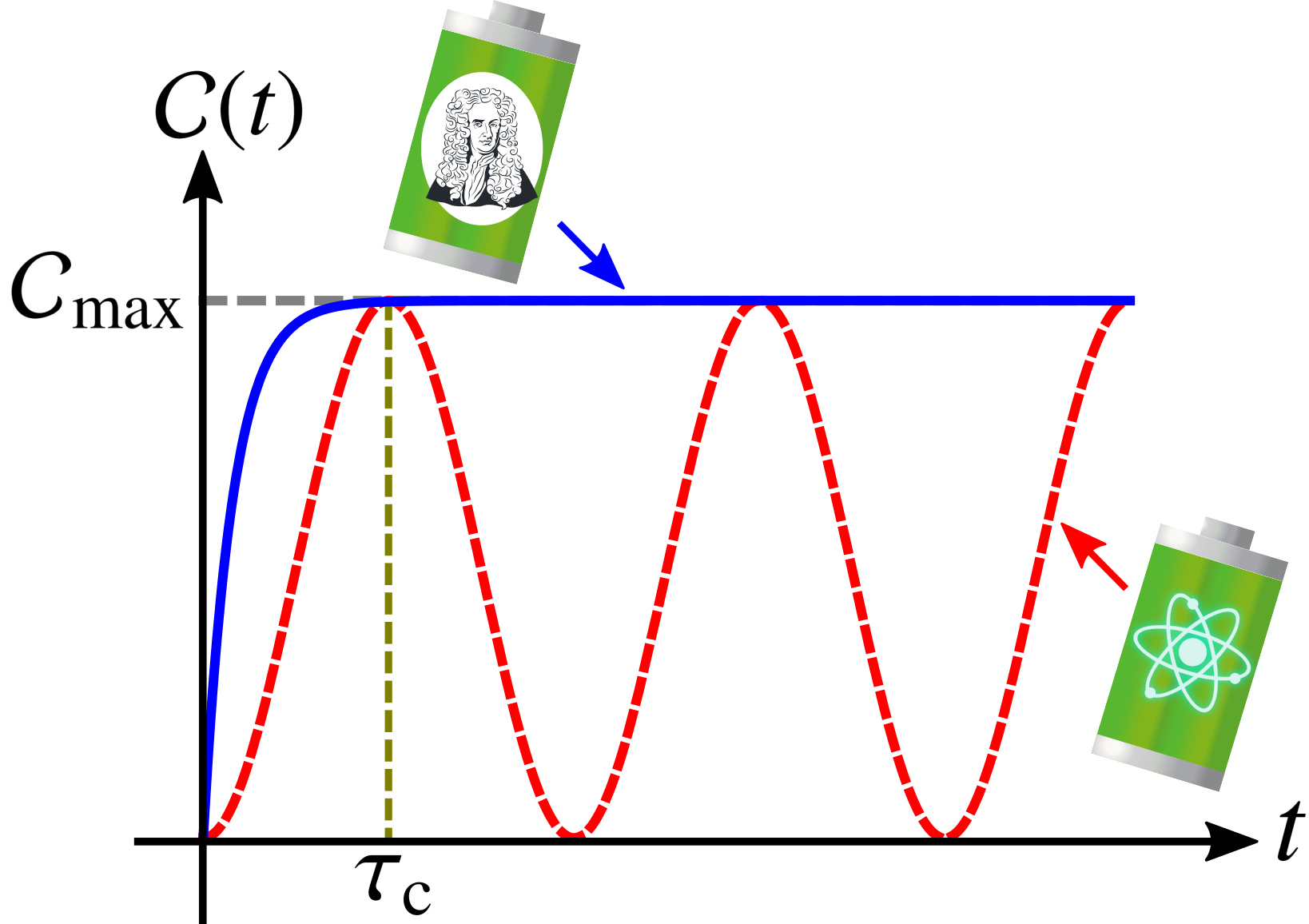}\includegraphics[width=0.5\columnwidth]{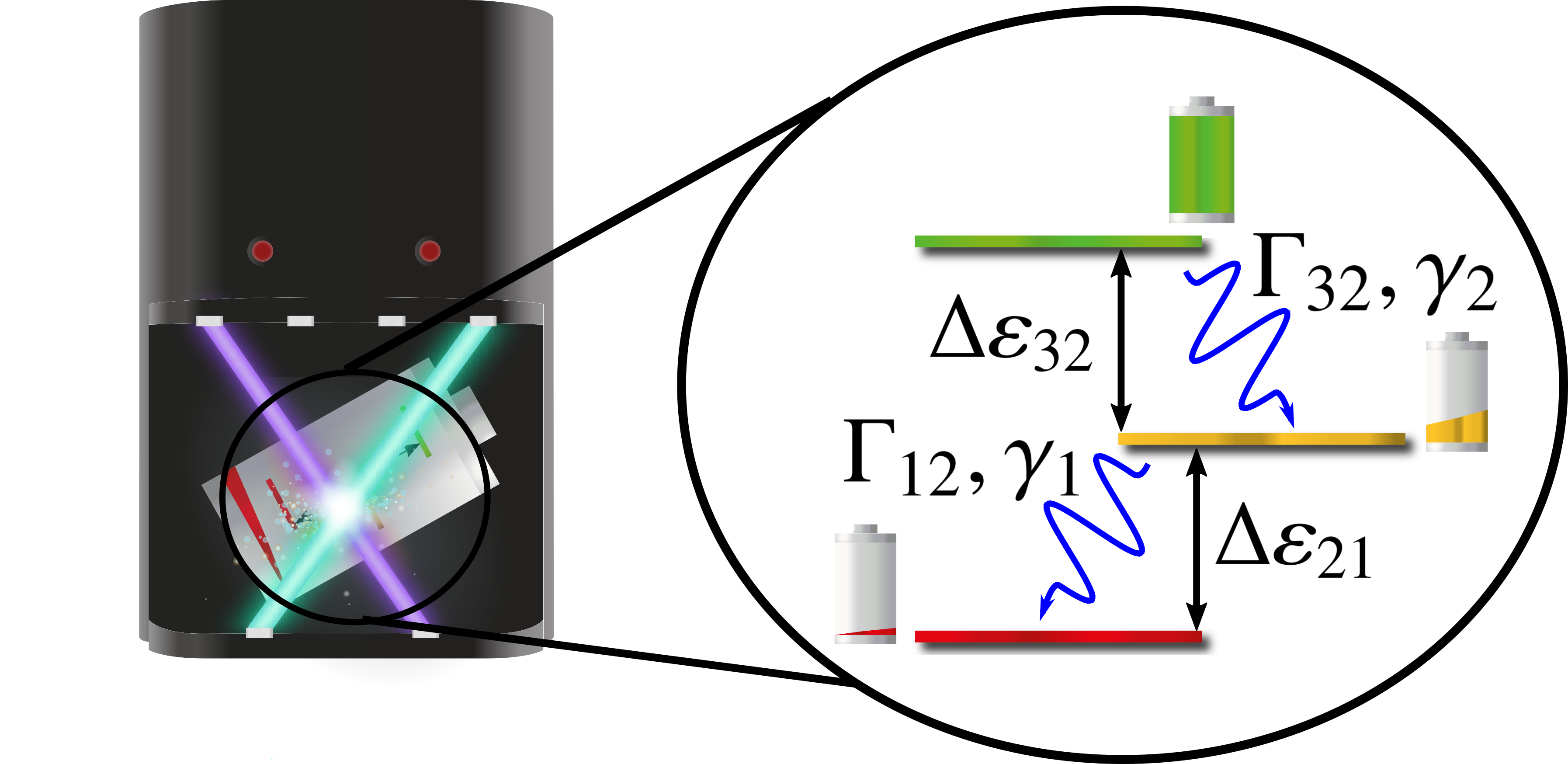}
\caption{(a) Schematic representation of the stored charge in classical and quantum batteries. While classical batteries present a charge stability after the charging time $\tau_{\text{c}}$, quantum batteries typically have oscillatory behavior. (b) Pictorial representation of a three-level quantum battery and its charger based on resonant fields. Each energy level represents a charging step in which the energy gap between them is $\Delta \varepsilon_{nm} \!=\! \varepsilon_{n} - \varepsilon_{m}$. In real physical settings, relaxation and dephasing could induce some non-unitary discharging by losses described by rates $\Gamma$ and $\gamma$, respectively.}
\label{FigScheme1}
\end{figure}

Quantum batteries have most commonly been proposed as an array of two-level systems in a number of different scenarios~\cite{Alicki:13, Ferraro:18,Le:18,Andolina:18-1,Andolina:18-2,Binder:15,PRE2019Batteries,PRL_Andolina,arXiv_Andolina}. In such settings when the battery is full, the charging field must be precisely switched off to avoid \textit{spontaneous discharging} due to coherent oscillations of the system that makes it bounce back and forth between charged and uncharged states~\cite{Andolina:18-1,Ferraro:18,Le:18,Andolina:18-2,PRE2019Batteries}, as schematically represented in Fig.~\ref{FigScheme1}(a). Therefore, in these systems successfully charging the quantum battery depends on the ability to decouple it from its ``charger", which typically are some external fields. The effect of the amplitude of the charging fields on the performance of the battery is two-fold: {\it (i)} the resulting oscillatory behavior is more pronounced if the fields are stronger, and {\it (ii)} a minimal time required to charge the battery, determined by the quantum speed limit for the evolution of the system, strongly depends on the strength of the external field~\cite{Campaioli:18}. While these conditions may not be particularly debilitating for certain realizations, it nevertheless leads us to ask whether we can design alternative schemes for charging a quantum battery that, in line with the classical counterpart, does not require disconnecting from its charger after the process is complete and furthermore that is robust to environmental effects.

In this work we provide such an alternative scheme. We consider a quantum battery possessing optimal characteristics for energy storage, regardless of the fields that act on it after it is fully charged, which also helps in designing a more robust battery against systematic errors. We propose a quantum battery using a three level quantum system (qutrit) that is externally driven in such a way that we have two different paths which connects the completely uncharged initial state to the fully charged state. We show that while one path still exhibits unwanted spontaneous discharging, as our battery is a three-level system we can exploit stimulated raman adiabatic passage (STIRAP)~\cite{Vitanov:17} to ensure a stable adiabatic quantum battery. Moreover, in order to provide a more realistic description, we consider the effects of dissipation and decoherence during the charging process on both the stored charge and the power of the charging process. Furthermore, we show that such high-dimensional quantum batteries provide a means to avert {\it self-discharging} by tuning the relative energy gaps. We provide an experimentally feasible scheme by discussing how our three-level quantum battery can be encoded and implemented in a superconducting transmon qutrit.

\section{Three-level quantum batteries}
A non-degenerate quantum battery is a $d$-level quantum system described by the Hamiltonian
\begin{eqnarray}
H_{0} = \sum_{n}^d \varepsilon_{n} \ket{\varepsilon_{n}}\bra{\varepsilon_{n}} \text{ , } \label{H0}
\end{eqnarray}
with $\varepsilon_{1} < \varepsilon_{2} < \dots < \varepsilon_{d}$. The battery is said to be in a {\it passive} state if no energy can be extracted from it through some cyclic process, e.g. the turning on of some potential $V(t)$, for a time $T$, that satisfies the boundary conditions $V(0)\!=\!V(T)\!=\!0$. Interestingly, an array of such batteries may not be passive with some work being available, but requiring some collective processing of the batteries~\cite{Binder:15, PRL2017Binder,Le:18,Ferraro:18}. A notable exception to this is thermal states which are said to be {\it completely passive} as even with such collective processing one cannot extract any work~\cite{Pusz:78}. {\it Active} states are those allow for work to be extracted through some cyclic process, with maximal amount of extractable work is called the ergotropy~\cite{Allahverdyan:04}. We remark that $d\!=\!2$ is a special case where all passive states are completely passive as any diagonal state in the energy eigenbasis is necessarily a thermal state at some temperature. In what follows we will be concerned with stable charging of a single three-level quantum battery, $d\!=\!3$. It is possible to generalize the proposed setting to systems with $d\!>\!3$ using chain STIRAP processes that connect the lowest lying energy eigenstate to the highest one~\cite{Vitanov:17}.

We will be interested in examining the energy stored in a quantum battery through some, possibly non-unitary, process. The energy at time $t$ is simply $E(t) = \trs{H_{0} \rho(t)}$ and if we assume our battery begins in the ground state, $\ket{\varepsilon_1}$, the ergotropy is the difference in energy between the final and initial battery states after the charging process is over,
\begin{eqnarray}
\Ccal (t) = E(t) - E_{\text{gs}} = \trs{H_{0} \rho(t)} - \varepsilon_{1} \text{,}
\end{eqnarray}
with $\Ccal_{\text{max}} = \hbar (\omega_{3}-\omega_{1})$ achievable via a process which transfers all the population from the initial ground state to the maximally excited state.

To drive the system and promote transitions between the energy levels we use auxiliary fields, which constitute a transitional Hamiltonian $H_{\text{t}}(t)$. In general, the Hamiltonian $H_{\text{t}}(t)$ depends on the structure of the system, but if we adequately order the basis $\{\ket{\varepsilon_{n}}\}$ the Hamiltonian $H_{\text{t}}(t)$ can be written in a general way as~\cite{Vitanov:17,Marangos:98,Fleischhauer:05,Bergmann:98}
\begin{equation}
H_{\text{t}}(t) = \hbar\Omega_{12}(t)e^{-i\omega_{12}t} \ket{\varepsilon_{1}}\bra{\varepsilon_{2}} + \hbar\Omega_{23}(t)e^{-i\omega_{23}t} \ket{\varepsilon_{2}}\bra{\varepsilon_{3}} + \text{h.c}  \text{ . } \label{Ht}
\end{equation}
In this case, the complete Hamiltonian which describes the dynamics of the system can be written as $H(t)\!=\!H_{0} + H_{\text{t}}(t)$. The Hamiltonian $H_{\text{t}}(t)$ develops the role of a \textit{quantum charger} for our three-level quantum battery as we need to couple our system to the external fields described by $H_{\text{t}}(t)$ in order to charge the battery. We sketch our proposal in Fig.~\ref{FigScheme1}(b). %Of course one could consider a driving Hamiltonian that directly couples the states $\ket{\varepsilon_{1}}$ and $\ket{\varepsilon_{3}}$, similar to the one in~\cite{Binder:15}.  Such a scenario is also a possible way to charge the battery, however, it will again bear the problem of unstability of the charging process, i.e. oscillations between charged and uncharged states, which we especially try to avoid.

While the bare Hamiltonian is important for dictating the amount of energy stored in the battery, the dynamics is driven by the interaction Hamiltonian $H_{\text{t}}(t)$. In fact, by considering the dynamics of the system in a general time-dependent interaction picture, the new Hamiltonian can be written as~\cite{Whaley:84}
\begin{eqnarray}
\dot{\rho}_{\text{int}}(t) = \frac{1}{i\hbar} [H_{\text{int}}(t),\rho_{\text{int}}(t)] \text{ . }
\end{eqnarray}
where $\rho_{\text{int}}(t) = e^{iH_{0}t} \rho(t) e^{-iH_{0}t}$ and
\begin{eqnarray}
H_{\text{int}}(t) = \hbar \Omega_{12}(t) \ket{\varepsilon_{1}}\bra{\varepsilon_{2}} + \hbar \Omega_{23}(t) \ket{\varepsilon_{2}}\bra{\varepsilon_{3}} + \text{h.c} \text{ , }
\end{eqnarray}
where we already assumed that both fields in Eq.~\eqref{Ht} are on resonance with the energy levels of the battery. Thus, it is possible to show that, in this new representation, we can get the population in each energy level from $P_{n} = \trs{\hat{P}_{n}\rho_{\text{int}}(t)} \!=\! \trs{\hat{P}_{n}\rho(t)}$, where $\hat{P}_{n}$ is the projector $\hat{P}_{n} \!=\! \ket{\varepsilon_{n}}\bra{\varepsilon_{n}}$. In addition, $\trs{H_{0}\rho_{\text{int}}(t)}\! =\! \trs{H_{0}\rho(t)}$, so that $\Ccal(t) \!=\! \trs{H_{0}\rho_{\text{int}}(t)}-E_{\text{gs}}$ can be obtained from the dynamics in the rotating frame. Therefore, we can consider the above equation in our study without loss of generality. We are interested in studying the charging procedure of our battery through an adiabatic dynamics in this new frame. By computing the set of eigenvectors of the new Hamiltonian $H_{\text{int}}(t)$ we find
\begin{subequations}
\label{EigenStates}
\begin{align}
\ket{E_{-}(t)} &= \frac{1}{\sqrt{2}} \left[
\frac{\Omega_{12}(t)}{\Delta(t)} \ket{\varepsilon_{1}} -
\ket{\varepsilon_{2}} + 
\frac{\Omega_{23}(t)}{\Delta(t)} \ket{\varepsilon_{3}}
\right] \\
\ket{E_{0}(t)} &= \frac{1}{\sqrt{2}} \left[
\frac{\Omega_{23}(t)}{\Delta(t)} \ket{\varepsilon_{1}} -
\frac{\Omega_{12}(t)}{\Delta(t)} \ket{\varepsilon_{3}}
\right] \label{E0}\\
\ket{E_{+}(t)} &= \frac{1}{\sqrt{2}} \left[
\frac{\Omega_{12}(t)}{\Delta(t)} \ket{\varepsilon_{1}} +
\ket{\varepsilon_{2}} + 
\frac{\Omega_{23}(t)}{\Delta(t)} \ket{\varepsilon_{3}} 
\right]\text{ , }
\end{align}
\end{subequations}
associated with eigenvalues $E_{\pm}(t) = \pm \hbar \Delta(t)$ and $E_{0}(t) =0$, where $\Delta^2(t) = \Omega_{12}^2(t)+\Omega_{23}^2(t)$. 

As mentioned, we will assume the process starts with the battery state $\ket{\psi(0)}\! =\! \ket{\varepsilon_{1}}$. This state can be written as a combination of different elements of Eqs.~\eqref{EigenStates}, depending on the initial values of the parameters $\Omega_{12}(0)$ and $\Omega_{23}(0)$. Therefore we can consider different charging protocols associated with distinct choices of the parameters $\Omega_{12}(t)$ and $\Omega_{23}(t)$, by adjusting how the external fields act on the system at the start of the evolution. We will show that while some protocols will lead to an unstable charged state, and therefore would require a carefully timed decoupling of the battery from the charging fields, by exploiting the Stimulated Raman Adiabatic Passage (STIRAP) technique, we can achieve a stable and robust charged state.

\subsection{Unstable charging}
From Eqs.~\eqref{EigenStates}, it is possible to show that the initial state $\ket{\psi(0)}$ can be written as a combination of the states $\ket{E_{-}(0)}$ and $\ket{E_{+}(0)}$ if we set $\Omega_{12}(0) \neq 0$ and $\Omega_{23}(0) = 0$. In fact, by considering this initial value we can show that
\begin{eqnarray}
\ket{\psi(0)} = \frac{\ket{E_{+}(0)} + \ket{E_{-}(0)}}{\sqrt{2}} = \ket{\varepsilon_{1}} \text{ . }
\end{eqnarray}
Allowing the system undergo adiabatic dynamics, we find the evolved state~\cite{Sarandy:04,Kato:50,Messiah:Book,Amin:09}
\begin{equation}
\ket{\psi^{\text{ad}}(t)} = \frac{1}{\sqrt{2}}\left[ e^{-\frac{i}{\hbar}\int_{0}^{t} E_{+}(t^{\prime})dt^{\prime}}\ket{E_{+}(t)} + e^{-\frac{i}{\hbar}\int_{0}^{t} E_{-}(t^{\prime})dt^{\prime}}\ket{E_{-}(t)} \right] \text{ , } \label{AdCuns}
\end{equation}
where we already used the parallel transport condition $\interpro{E_{n}(t)}{\dot{E}_{n}(t)}\!=\!0$, for all $n$. Thus, we write
\begin{equation}
\ket{\psi^{\text{ad}}(t)} = \frac{\cos\Phi(t)}{\Delta(t)}\Big(\Omega_{12}(t) \ket{\varepsilon_{1}} + \Omega_{23}(t) \ket{\varepsilon_{3}} \Big) - i \sin\Phi(t) \ket{\varepsilon_{2}},
\end{equation}
where $\Phi(t)\!=\! \int_{0}^{t} \Delta(t)d\xi$. Therefore, one finds the ergotropy $\Ccal(t)\! =\! \bra{\psi^{\text{ad}}(t)}H_{0}\ket{\psi^{\text{ad}}(t)} - \bra{\varepsilon_{1}}H_{0}\ket{\varepsilon_{1}}$ as
\begin{align}
\Ccal(t) &= \hbar \frac{\cos^2\Phi(t)}{\Delta^2(t)}\left[ \omega_{1} \Omega^2_{12}(t) + \omega_{3}\Omega^2_{23}(t)\right] + \hbar \omega_{2} \sin^2\Phi(t) \nonumber \\
& - \hbar\omega_{1} \label{Cuns} \text{ . }
\end{align}

To achieve maximal ergotropy firstly we must fix the final values for the parameters $\Omega_{12}(t)$ and $\Omega_{32}(t)$ at some cutoff time $\tau_c$ in order to get $\Omega_{12}(\tau_{\text{c}})\!=\!0$ and $\Omega_{32}(\tau_{\text{c}})\!\neq\!0$. This involves particular initial and final conditions on the parameters $\Omega_{12}(t)$ and $\Omega_{32}(t)$ to fully charge the battery. Secondly, the instant in which the system achieves the full charge is when $\cos\Phi(\tau_{\text{c}}) = 1$. Under these constraints, we achieve maximum ergotropy, $\Ccal(\tau_{\text{c}})\!=\! \Ccal_{\text{max}}$. However, from Eq.~\eqref{Cuns} one can see that for $t \!>\! \tau_{\text{c}}$ the battery charge cannot be kept at its maximum value, and rather it will continue to oscillate between fully charged and fully dissipated states due to the action of the fields. We describe a protocol which leads to this situation as an unstable battery charging process. In addition, the function $\Phi(t)$ depends on the integration from $0$ to some instant $t\!>\!\tau_{\text{c}}$, the sine and cosine functions could become highly oscillating, such that that after $t\!>\!\tau_{\text{c}}$ we can have many maximum and minimum values for the ergotropy. We understand this as follows: in an adiabatic regime of the charging process, there is an intrinsic discharging process due to the relative quantal phases in Eq.~\eqref{AdCuns}. The adiabatic phase associated with different adiabatic paths (eigenstates), promote destructive and constructive superpositions of the components $\ket{\varepsilon_{3}}$ of the states $\ket{E_{+}(t)}$ and $\ket{E_{-}(t)}$. Consequently, we observe the natural discharging as a process due to destructive interference from $\ket{\varepsilon_{3}}$. Thus, a charging strategy that begins the adiabatic evolution with $\Omega_{12}(0)\!\neq \!0$ and $\Omega_{32}(0) \!=\! 0$ does not lead to a stable and robust quantum battery. We remark that the above result is not a particular feature of adiabatic charging process. Actually, this spontaneous discharging is an intrinsic characteristic of different systems where the oscillatory behavior of the quantal phases promotes some (partial) destructive interference as obtained in Eq.~\eqref{Cuns}.

\subsection{Stable charging via STIRAP}
An alternative strategy for our quantum battery is through the eigenstate $\ket{E_{0}(t)}$, the so-called \textit{dark state}~\cite{Fleischhauer:05}. In order to follow this path, we need to set the initial values of the parameters $\Omega_{12}(0)\! =\! 0$ and $\Omega_{32}(0)\!\neq \!0$. Thus,
\begin{equation}
\ket{\psi(0)} = \ket{E_{0}(0)} = \ket{\varepsilon_{1}} \text{ , } \label{AdCstb}
\end{equation}
By letting the system undergo adiabatic dynamics, the evolved state becomes
\begin{equation}
\ket{\psi^{\text{ad}}(t)} = \ket{E_{0}(t)} = \frac{1}{\sqrt{2}} \left[
\frac{\Omega_{23}(t)}{\Delta(t)} \ket{\varepsilon_{1}} -
\frac{\Omega_{12}(t)}{\Delta(t)} \ket{\varepsilon_{3}}
\right] 
\end{equation}
with no quantal phase accompanying the evolution, because the adiabatic phase is null, once we have $E_{0}(t)\!\!=\!\!0$ and $\interpro{E_{0}(t)}{\dot{E}_{0}(t)}\! =\! 0$. The ergotropy is then
\begin{eqnarray}
\Ccal(t) = \hbar \frac{\omega_{3}\Omega_{12}^2(t) + \omega_{1}\Omega_{23}^2(t)}{\Delta^2(t)} - \hbar\omega_{1} \label{Cstb} \text{ , }
\end{eqnarray}
which achieves its maximumal value when $\Omega_{12}(\tau_{\text{c}})\!\neq\! 0$ and $\Omega_{23}(\tau_{\text{c}}) \!=\! 0$, without any assumption about the value of $\tau_{\text{c}}$, in stark contrast to the unstable charging process. Clearly to get a fully charged battery both initial and final conditions on the parameters $\Omega_{12}(t)$ and $\Omega_{32}(t)$ are required. However, by exploiting the STIRAP protocol we can avoid the oscillatory behavior otherwise present due to accumulated quantal phases.

A second important physical lesson of these results are associated with the intrinsic characteristics of dark states. Unlike the other eigenstates of the Hamiltonian driving the system, the dark state does not allow population inversion even when we put the fields on resonance with the system. This property allows us to design a robust battery that does not suffer from spontaneous discharging if the control fields are not switched off after the charging process. Thus, the emergence of the dark state further highlights the relevance of three-level (or $N$-level) systems over the more commonly considered two-level qubits in designing stable quantum batteries~\cite{CampbellBatteries}.

\section{Relaxation and Dephasing Effects}
So far we have focused on an idealized setting where our quantum battery does not suffer any environmentally induced spoiling effects. In this section we consider the performance and stability of our quantum battery when the most relevant environmental effects are taken into consideration (see Refs.~\cite{PRA_Ivanov,Vitanov:17} for other studies exploring decoherence effects on STIRAP protocols). In particular we will consider a dynamics governed by a Lindblad master equation~\cite{Lindblad:76} which takes into account both relaxation and dephasing phenomena, corresponding to the most natural non-unitary effects in superconducting circuits~\cite{Li:11,Martinis:03,JPCS_Li}, which we propose as a natural platform to realize our battery as we will elaborate on in Sec.~\ref{sec:transmon}. The dynamics of the system is given by
\begin{eqnarray}
\dot{\rho}_{\text{int}}(t) = \frac{1}{i\hbar} [H_{\text{int}}(t),\rho_{\text{int}}(t)] + \Lcal_{\text{rel}}[\rho_{\text{int}}(t)]+ \Lcal_{\text{dep}}[\rho_{\text{int}}(t)] \text{ , } \label{LindEq}
\end{eqnarray}
where the superoperators $\Lcal_{\text{rel}}[\bullet]$ and $\Lcal_{\text{dep}}[\bullet]$ describe the relaxation and dephasing phenomena, respectively, and can be written as
\begin{subequations}
\begin{align}
\Lcal_{\text{rel}}[\bullet] &= \sum_{k\neq j}\Gamma_{kj} \left[\sigma_{kj}\bullet\sigma_{jk} - \frac{1}{2}\{\sigma_{kk},\bullet\} \right] \text{ , } \\
\Lcal_{\text{dep}}[\bullet] &= \sum_{j=2,3}\gamma_{j} \left[\sigma_{jj}\bullet\sigma_{jj} - \frac{1}{2}\{\sigma_{jj},\bullet\} \right] \text{ , }
\end{align}
\label{RelTerm}
\end{subequations}
where $\sigma_{kj}\!=\!\ket{\varepsilon_{k}}\bra{\varepsilon_{j}}$ and $\Gamma_{kj}\!=\!\Gamma_{jk}$. Building on the general definitions we have introduced in Eqs.~\eqref{RelTerm}, we would like to clarify two important points on the characteristics of noise we consider in the rest of this work. First, the relaxation processes we consider are only the sequential decays, meaning, $\ket{\varepsilon_{3}}\! \rightarrow \!\ket{\varepsilon_{2}}$ and $\ket{\varepsilon_{2}} \!\rightarrow\! \ket{\varepsilon_{1}}$ characterized by the rates $\Gamma_{32}$ and $\Gamma_{21}$, respectively. We do not take into account the nonsequential decay mechanism which is responsible from inducing transitions like $\ket{\varepsilon_{3}} \!\rightarrow\! \ket{\varepsilon_{1}}$, since the rate associated with such a process, $\Gamma_{31}$, is an order of magnitude smaller for transmon qubits~\cite{Peterer:15} (as we shall discuss later). Second, $\gamma_2$ and $\gamma_3$ determine the rates at which the superpositions between $\ket{\varepsilon_{1}}$ and $\ket{\varepsilon_{2}}$, and $\ket{\varepsilon_{1}}$ and $\ket{\varepsilon_{3}}$ are suppressed, respectively. Together, they also contribute to the dephasing of superpositions between $\ket{\varepsilon_{2}}$ and $\ket{\varepsilon_{3}}$. However, due to the nature of the STIRAP protocol with the dark state, the only dephasing rate that has an impact on the charging protocol is $\gamma_3$, since the state $\ket{\varepsilon_{2}}$ is never populated during the process. Clearly, there are a number of timescales and relevant noise parameters to fix in order to quantitatively assess the effectiveness of our protocol. In what follows, we will focus on those parameter ranges most relevant for transmon qubits, which provide a promising candidate architecture. Nevertheless, we expect the qualitative behavior discussed to hold in other relevant settings.

\subsection{Stable charging under dissipation and decoherence}

We begin examining the effect that environmental spoiling mechanisms have on the charging process itself. To this end we consider $\Omega_{12}(t) \!=\! \Omega_{0} f(t)$ and $\Omega_{23}(t) \!=\! \Omega_{0} [1-f(t)]$, where $f(t)$ is a function which satisfies $f(0) = 0$ and $f(\tau) = 1$, such that the boundary conditions on $\Omega_{12}(t)$ and $\Omega_{23}(t)$ are satisfied and we realise the stable charging via STIRAP. We can readily examine the behavior of the ergotropy as a function of the dimensionless parameter $\Omega_{0}\tau$. In addition to the ergotropy, equally important is assessing the charging power of quantum batteries~\cite{Campaioli:18}, which we define as 
\begin{eqnarray}
\Pcal(\tau) = \frac{\Ccal(\tau)}{\tau} \text{ , } \label{CP}
\end{eqnarray}
where $\Ccal(\tau)$ is the amount of energy transferred to the battery from external fields during the time interval $\tau$. In order to make a meaningful comparison, we rescale $\Pcal$ with the maximal attainable power $\Pcal_\text{max}$. As argued by Binder {\it et al}, it is physically reasonable to bound the amount of energy available for a given charging protocol~\cite{Binder:15}. The most efficient charging process therefore corresponds to one which needs only enough energy to fully charge the battery, in our case $\hbar(\omega_3 - \omega_1)$. We can then exploit the quantum speed limit~\cite{DeffnerReview} to determine the minimum time, $\tau_\text{QSL}$, needed for some time-independent process to charge the battery and thus corresponds to the most powerful charging obtainable, under this energy constraint~\cite{Binder:15}. Thus,  $\Pcal_\text{max}\!=\!\pi/(2\hbar(\omega_3 - \omega_1))$. We fix the functional form of $f(t)$ to be a simple linear ramp, $f(t) \!=\! t / \tau$. Naturally, one could consider any other ramp that satisfies the boundary conditions, however, as STIRAP is an adiabatic protocol, the means by which one manipulates the system is of little consequence. While from one ramp to another some qualitative differences may emerge in the behavior as one approaches the adiabatic regime, the quantitative features outlined in what follows persist.

\begin{figure}[t]
{\bf (a)} \\
%\input{ergotropy.plt}\\
%\vspace{0.25cm}
\includegraphics[width=0.92\columnwidth]{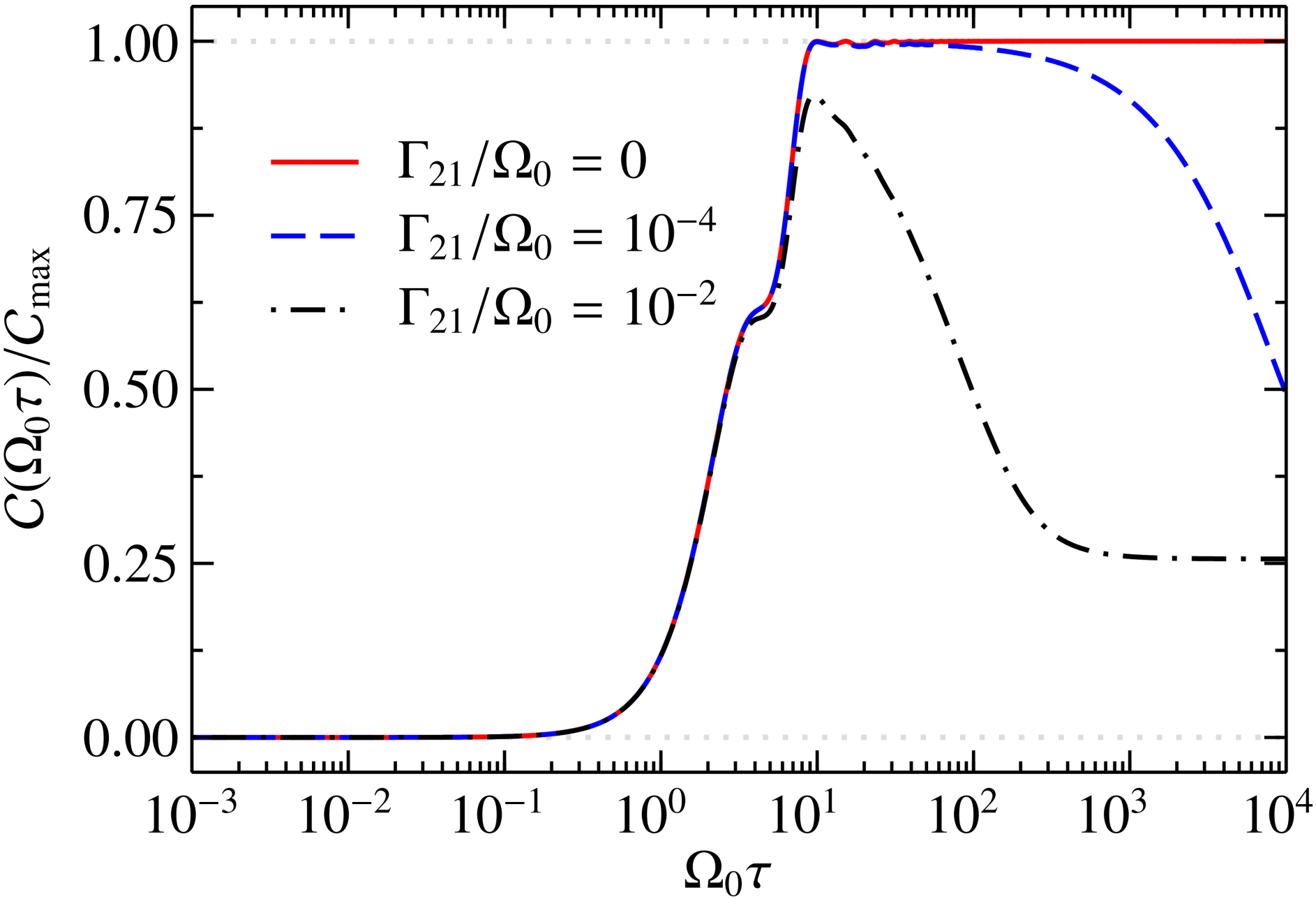} \\
\vspace{0.25cm}
{\bf (b)} \\
\includegraphics[width=0.92\columnwidth]{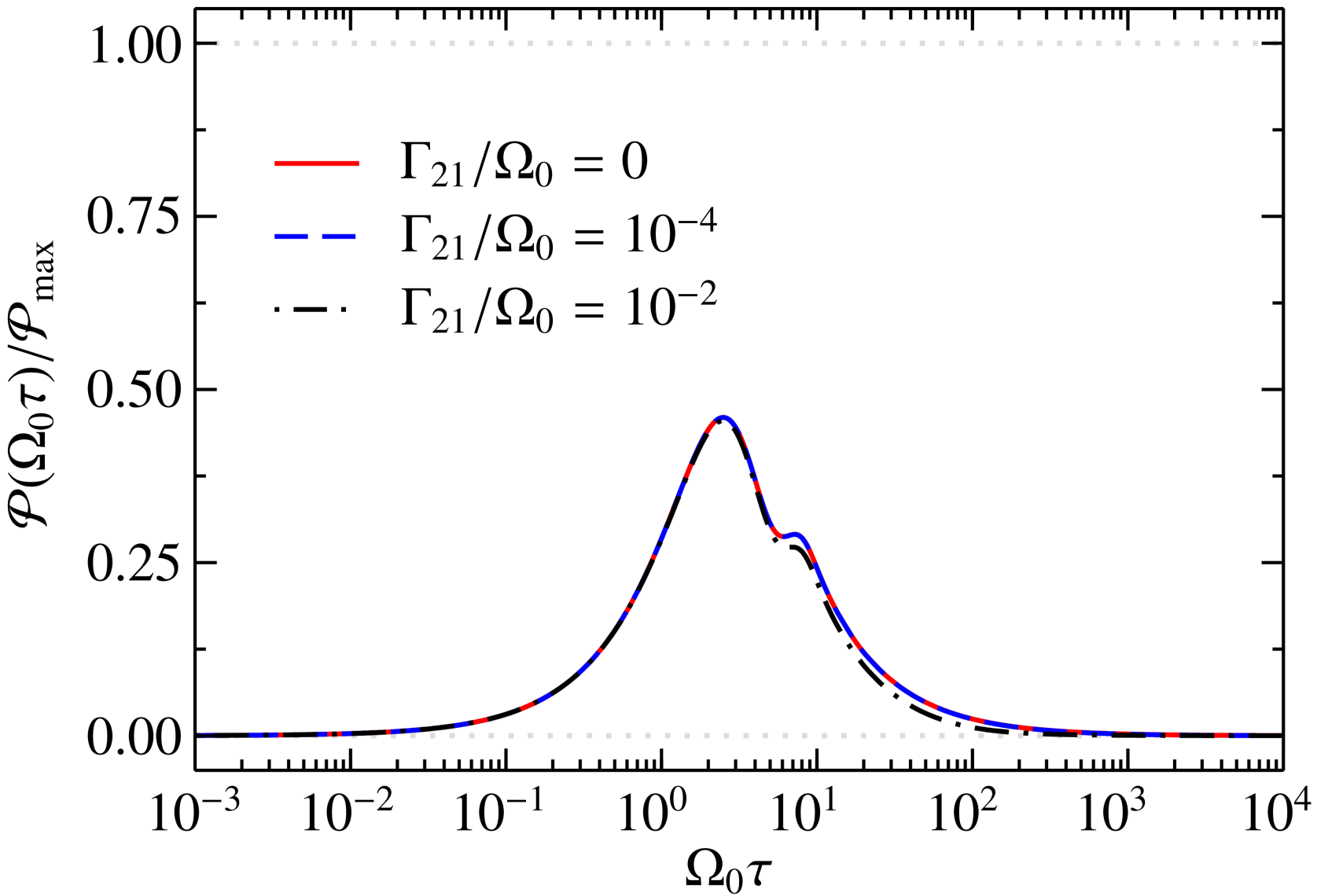} \\
\caption{(a) Ergotropy and (b) power for a linear ramp including the effects of both relaxation and dephasing. We chose the energy spectrum of our three-level system as $\omega_1\!=\!0$, $\omega_2\!=\!1$ and $\omega_3\!=\!1.95$ in order to account for the anharmonicity of the energy gaps in transmon qubits, resulting in $\Ccal_{\text{max}}\!=\!1.95\hbar$. We set the rates characterizing the noise as $\Gamma_{32}\!=\!2\Gamma_{21}$, $\gamma_2\!=\!\Gamma_{21}$, and $\gamma_3\!=\!\Gamma_{32}\!=\!2\Gamma_{21}$, again to match the state-of-the-art parameters measured for transmon qubits~\cite{Peterer:15}.  }
\label{FigRel}
\end{figure}

In Fig.~\ref{FigRel}(a) we show the ergotropy as a function of $\tau$ for several values of decoherence and dissipation which we specify in detail in the caption and are inline with the transmon implementation we propose in the following section. The topmost solid curve corresponds to no decoherence and we see a fast evolution gives a vanishingly small ergotropy as these timescales are far from the adiabatic limit, therefore the STIRAP protocol is ineffective and no population inversion can be observed. As we increase $\tau$, in line with the adiabatic theorem~\cite{Sarandy:05-1,Amin:09,Hu-18-b}, the maximum ergotropy grows and we achieve a fully charged state when the STIRAP protocol is faithfully implemented. We clearly see that in the case of no decoherence the charged state is perfectly stable for $\tau\!\gtrsim\!10/\Omega_{0}$. Conversely, the ergotropy is affected when the decoherence effects become more significant. For small values of decoherence (blue, dashed curve) the STIRAP protocol is quite robust and only becomes significantly adversely affected when the times scales are an order of magnitude slower than strictly necessary. As the environmental effects are increased we find that achieving a fully charged battery is not possible, however, we can identify a range of values for $\tau$ for which we get the optimal stored charge, cf. the peak of the green, dotted curve in Fig.~\ref{FigRel}. Thus, a given $\tau$ sets the speed of the adiabatic evolution and we can see that an optimality criterion between total evolution time and decoherence effects appears.

The rescaled power is shown in Fig.~\ref{FigRel}(b), which is only weakly affected for reasonable environmental parameters. Naturally, for fast protocols where the battery fails to charge the resulting power is negligible. As $\tau$ increases the charging power also increases until it reaches a maximum of $\!\sim\!0.5\Pcal_\text{max}$. The discrepancy between the maximum obtained power and $\Pcal_\text{max}$ is due to the fact the latter is based on the quantum speed limit time, which is typically much shorter than the adiabatic timescales required for our protocol to be effective. However, it is interesting to note that the maximum power does not correspond to when the battery is fully charged. By comparing Figs~\ref{FigRel}(a) and (b) we see that, for all the considered noise values, the maximum ergotropy is achieved for $\tau\!\sim\!10/\Omega_{0}$, which corresponds to $\Pcal\!\sim\!0.25\Pcal_\text{max}$. Thus we find that there is a trade-off between the maximum achievable ergotropy and the power when stably charging a quantum battery via STIRAP. A promising method to boost the power of our protocol would be to employ so-called shortcuts-to-adiabaticity~\cite{STAreview}. However, these techniques invariably come at the cost of some additional resources which will affect the resulting efficiency and power, but nevertheless may prove useful to ensure both fast and stable quantum batteries.

It is important to stress that the results shown in Fig.~\ref{FigRel} do not take into account decay transitions between $\ket{\varepsilon_{3}} \!\rightarrow\! \ket{\varepsilon_{1}}$. This assumption is justified since, in case where no noise mechanism acts on our system, the adiabatic behavior is achieved for $\tau\Omega_{0} \sim 10$. Notice that the highest decay rate we consider is $\Gamma_{21}/\Omega_{0} \leq 10^{-2}$, which allows us to write $(\Gamma_{21}\tau/\tau\Omega_{0})\leq 10^{-2}$ or $\Gamma_{21}\tau \leq 10^{-2}\tau\Omega_{0}$. Replacing the parameters with their values in the adiabatic limit we find $\Gamma_{21}\tau \!\lesssim \!10^{-2}10 = 10^{-1}$. Thus, $\Gamma_{21}$ is indeed relevant to our discussion. As experimentally shown~\cite{Peterer:15}, the timescale $\tau_{31}$ for the process $\ket{\varepsilon_{3}} \!\rightarrow\! \ket{\varepsilon_{1}}$ is $\tau_{31} \! \sim \! 10^2 \tau_{21}$, where $\tau_{21}$ is the timescale of the process $\ket{\varepsilon_{2}} \!\rightarrow\! \ket{\varepsilon_{1}}$. Therefore we can write $\Gamma_{31} \sim 10^{-2} \Gamma_{21}$. From this we have that $\Gamma_{31}\tau \sim 10^{-2} \Gamma_{21}\tau \lesssim 10^{-3}$, which allows us to conclude that the relaxation due to non-sequential rates, i.e. $\Gamma_{31}$, are negligible for the adiabatic time scales considered in this work.

\begin{figure}[t]
	\centering
	\includegraphics[width=0.93\columnwidth]{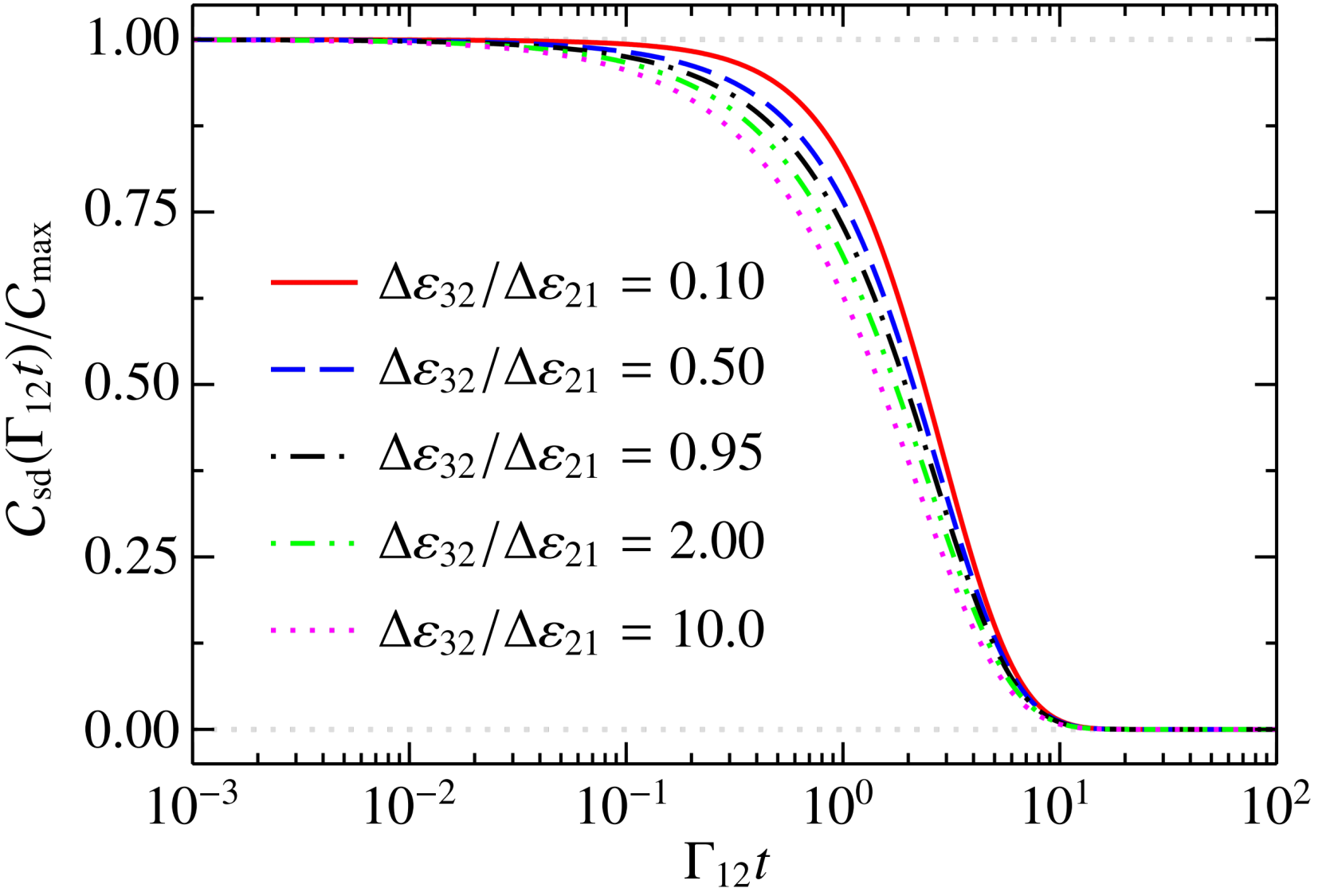}
	\caption{Ergotropy during a self-discharging process of our battery as function of $\Gamma_{21}t$ for different choices of the energy gaps, and the rest of the parameters remain as in Fig.~\ref{FigRel}. Note that the black dash-dotted line with $\delta\varepsilon_{32}/\delta\varepsilon_{21}\!=\!0.95$ corresponds to the energy gap configuration considered in Fig.~\ref{FigRel} and represents the case most relevant to a transmon qubit implementation.}
	\label{SelfDischFig}
\end{figure}
\subsection{Self-discharging of a quantum battery}
Left to their own devices, classical batteries are known to \textit{self-discharge}, a natural process associated with chemical reactions which reduce the stored charge even when the battery is not coupled to some device~\cite{Wu:00,Galushkin:12,Zhu:14,Shinyama:06}. Here we consider the same phenomena for our quantum battery as a natural process due to the relaxation effects on the system. Therefore, the initial state will be the charged state $\rho_{\text{c}}$ so that the ergotropy is
\begin{eqnarray}
\Ccal_{0} = \trs{\rho_{\text{c}}H_{0}} - \varepsilon_{1} = \sum\nolimits_{n}\varepsilon_{n}\varrho^{n}_{\text{c}} - \varepsilon_{1} \text{ , }
\end{eqnarray}
where $\varrho^{n}_{\text{c}}$ denotes the $n$-th diagonal element of $\rho_{\text{c}}$ corresponding to the population of the $n$-th energy level. We remark that while we will consider fully charged states, the proceeding results remain qualitatively unaffected for partially charged states. In the same way, we can write the instantaneous time-dependent ergotropy in the self-discharging process as
\begin{eqnarray}
\Ccal(t) = \trs{\rho(t)H_{0}} - \varepsilon_{1} = \sum\nolimits_{n}\varepsilon_{n}\varrho^{n}(t) - \varepsilon_{1}\text{ , }
\end{eqnarray}
with $\varrho^{n}(t)$ being the population at time $t$. Unlike the previous section, here we do not consider the effect of dephasing as the charged state is already diagonal in the energy eigenbasis. It is reasonable to assume that during the self-discharging process the quantum battery is no longer coupled to the external charging fields such that its dynamics is given by 
\begin{eqnarray}
\dot{\rho}(t) = \frac{1}{i\hbar} [H_{0},\rho(t)] + \Lcal_{\text{rel}}[\rho(t)] \text{ , } \label{LindEqSD}
\end{eqnarray}
with $\Lcal_{\text{rel}}[\bullet]$ given by Eq.~\eqref{RelTerm}. Since $\Ccal(t)$ depends only on the elements $\varrho^{nn}(t)$, the problem of finding $\Ccal(t)$ reduces to the task of solving the equations for a simple dissipative process
\begin{align}
\dot{\varrho}^{2}(t) &= -\Gamma_{21}\varrho^{2}(t) + \Gamma_{32}\varrho^{3}(t) \\
\dot{\varrho}^{3}(t) &= - \Gamma_{32}\varrho^{3}(t) \text{ , }
\end{align}
whose solution is given by (See Appendix~\ref{LindEqSDSol} for details)
\begin{align}
\varrho^{2}(t) &= \frac{\Gamma_{32}e^{-t\Gamma_{32}} - \Gamma_{32} e^{-t\Gamma_{21}}}{\Gamma_{21}-\Gamma_{32}} \\
\varrho^{3}(t) &= e^{-t\Gamma_{32}} \text{ . }
\end{align}
where we used the initial condition for a fully charged battery ($\varrho^{1}_{\text{c}}=\varrho^{2}_{\text{c}}=0$ and $\varrho^{3}_{\text{c}}=1$). Using the relation $\varrho^{1}(t) + \varrho^{2}(t) +\varrho^{3}(t) =1$, which is valid for every $t$, so we find
\begin{align}
\Ccal(t) &=  
\frac{e^{-t\Gamma_{32}}\left( \Gamma_{21}\Delta_{31} - \Gamma_{32}\Delta_{32}  \right) - e^{-t\Gamma_{21}}\Delta_{21}
	\Gamma_{32}}{\Gamma_{21}-\Gamma_{32}} \text{ , } \label{Csd}
\end{align}
where $\Delta_{mn} \!=\! \varepsilon_{m}-\varepsilon_{n}$ are the gaps between the energy levels of the system. In general the damping rates for the different energy gaps can be different, such that one finds that $\Gamma_{21} \!\neq\! \Gamma_{31}$. This means that the ergotropy of a three-level quantum battery is not dictated by a single exponential decay. Such a behavior is characteristic of classical supercapacitors as theoretically studied in Ref.~\cite{Kowal:11} for three different types of commercially available supercapacitors and experimentally verified in a carbon-based supercapacitors with organic electrolytes~\cite{Ricketts:00}. While Eq.~\eqref{Csd} accounts for the expected exponential decay, it nevertheless reveals that such effects can be tuned by modifying the internal structure of the battery. In Fig.~\ref{SelfDischFig} we see that by manipulating the relative energy gaps we can realize longer-lived stable quantum batteries. This is at variance with two-level systems where similar effects cannot be manipulated due to the presence of only a single energy splitting.

\section{Three-level superconducting transmon quantum battery}\label{sec:transmon}

\begin{figure}
	\centering
	\subfloat[Superconducting circuit]{\includegraphics[scale=1.9]{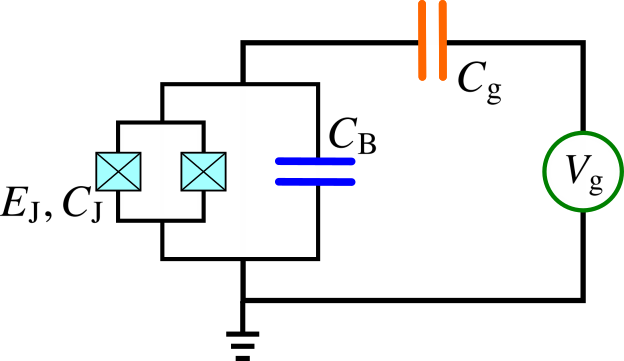}\label{TransmFig1}} \quad
	\subfloat[Transmon energy levels]{\includegraphics[scale=1.9]{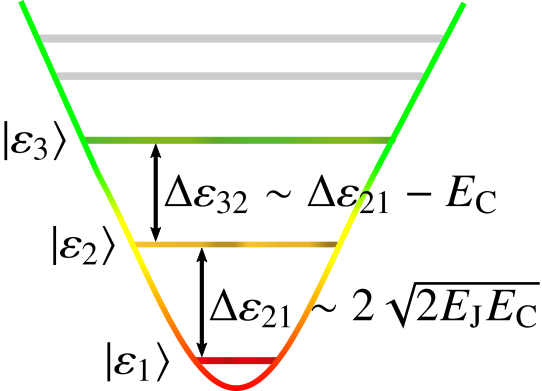}\label{TransmFig2}}
	\caption{(a) The sketch of a superconducting transmon qubit circuit, where a Josephson junction of capacitance $C_{\text{J}}$ is shunted by a large capacitance $C_{\text{B}}$. (b) Energy level structure of the transmon qutrit, where the maximum stored energy is given by means of the energies $E_{\text{J}}$ and $E_{\text{C}}$ as $\Ccal_{\text{max}} = \varepsilon_{3} - \varepsilon_{1} = \Delta\varepsilon_{32} + \Delta\varepsilon_{21} = 4\sqrt{2 E_{\text{J}}E_{\text{C}}} - E_{\text{C}}$.}
	\label{TransmFig}
\end{figure}

The three-level battery introduced here can be implemented in several physical systems in which we can encode a ladder three-level systems, like trapped ion systems and superconducting circuit QED system~\cite{devoret2012,wendin2017,gu2017}, for example. Here we propose that superconducting transmon qubits are particularly suitable candidates~\cite{Koch:07,you2007,Schreier08,barends2013}, the ladder-type three-level system is schematically presented in Fig.~\ref{TransmFig}. These qubits are fabricated (typically planar) chips and consist of two Josephson junctions, with capacitance $C_{\text{J}}$ and energy $E_{\text{J}}$, that are shunted by a large capacitor with capacitance $C_{\text{B}}$. While the quantized circuit corresponding to a standard $LC$ circuit (capacitance-inductance) will result in a harmonic oscillator, the Josephson junction functions as a non-linear inductor and distorts the spectrum of the oscillator away from the equally-spaced one. The great success of the transmon qubit derives from its large ratio of Josephson to capacitative energy $E_{\text{J}}/E_{\text{C}}$, where $E_{\text{C}} \!=\! e^2 /2C$ with $C \!=\! C_{\text{J}} +C_{\text{B}} + C_{\text{g}}$. As discussed by Koch {\it et al.}~\cite{Koch:07}, a large $E_{\text{J}}/E_{\text{C}}$ renders the system very insensitive to charge noise, hence enhancing the lifetime. There is a catch however: the anisotropy in the spectrum also scales with $E_{\text{J}}/E_{\text{C}}$ and goes down with increasing ratio. Recall that the anisotropy needs to be significant to ensure that we are away from the equally-spaced case and can address individual levels to produce well-defined qubits. Fortunately, the anisotropy scales as a power law, while the noise sensitivity depends exponentially on this ratio. Hence, one may find a ``sweet spot" with good anisotropy and long lifetimes. In practice, one typically aims for $E_{\text{J}}/E_{\text{C}}\!\sim\! 80-100$~\cite{Koch:07}. Typical energy level splittings in transmon qubits are of the order of 10~GHz, while the anisotropies are of the order of 100~MHz, and while this is much smaller than the splitting, modern microwave techniques are more than adequate to address such levels~\cite{gu2017}. 

Here we are interested in a transmon using its three lowest levels as active quantum states. These higher-dimensional local Hilbert spaces have been proposed as a way to simplify quantum logicoperations \cite{rungta2000,zhou2003,ralph2007} and experimental realization of important gates such as the Toffoli gate has been achieved using photons~\cite{lanyon2008}. While higher levels in transmons may be the source of unwanted leakage that must be minimized~\cite{stig2018,loft2018}, they may also serve useful purposes, such as in the coupling of cavity and qubits~\cite{wallraff2004,majer2007} to achieve effective $ZZ$ coupling terms from avoided crossings with higher levels in the spectrum~\cite{dicarlo2010}. More direct addressing of the higher levels of single transmon qubits has also been discussed \cite{Peterer:15,Liu:16}. Furthermore, a combination of two- and three-level systems may be used to do more effective quantum operations~\cite{bakkegaard2018} or quantum simulation~\cite{rafael2018}. 

In the case of the quantum battery, we need a system with three levels that are tunable and controllable. In a superconducting transmon qubit design, one may realize a three-level qutrit by adding flux and drive lines. In practice, this is done by adding a drive at node fluxes in the circuits~\cite{gu2017}. An example could be a sinusoidal drive at two nodal points that will induce a time-dependent driving term on the effective qubit/qutrit degrees of freedom that is similar to a dipole coupling of an electromagnetic field to an atom. This induces an ac-stark shift of the levels and establishes a set of dressed states. Using an appropriate drive line on the circuit, one can tune the levels, as well as drive population between them. Hence, a three-level system appropriate for the quantum battery can be realized using a superconducting transmon (with or without a cavity) with applied driving. We note that a qubit-qutrit combination~\cite{bakkegaard2018}, may be a good setup for not only implementing the battery, but also for probing its properties in a fully quantum manner by its coupling to a qubit system that can be read out.

\section{Conclusions}
We have shown that stable adiabatic quantum batteries are achievable for three-level systems. We employ stimulated raman adiabatic passage which allows one to bypass the undesired spontaneous discharging due to imprecise control on the fields that occur if the charging process couples directly only two levels of the battery, e.g. the ground and maximally excited states. Our protocol allows for the design of batteries that are robust to intrinsic errors in real physical scenarios concerning unknown delays in turning off the charging fields. While (effective) qubit batteries require careful manipulation of the charging fields, our three-dimensional quantum battery is able to exploit the STIRAP protocol to ensure a robust and stable charge. We explicitly consider the effects of the most relevant sources of noise and have shown that even for moderate values of decoherence and dissipation, our adiabatic quantum battery is quite robust. For more severe environmental effects we have shown that an optimal time emerges that dictates the maximal achievable ergotropy. Furthermore, we have established that self-discharging of high-dimensional quantum batteries can be mitigated by tuning the relative energy gaps. We finally proposed that superconducting transmon qubits provide a promising implementation for adiabatic quantum batteries. Our results show that clear advantages can be gained by exploiting higher-dimensional quantum systems. As such we expect that extending our analysis to consider arrays of high-dimensional quantum batteries, and the role of entanglement in the collective charging process, will be of significant interest~\cite{Alicki:13, Binder:15, PRL2013Huber}. 

\acknowledgments
ACS acknowledges financial support from the Brazilian agencies Conselho Nacional de Desenvolvimento Científico e Tecnológico (CNPq), Brazilian National Institute of Science and Technology for Quantum Information (INCT-IQ) and the Coordena\c{c}\~ao de Aperfei\c{c}oamento de Pessoal de N\'{\i}vel Superior - Brasil (CAPES) (Finance Code 001).
%%%%%%%%%%%%%%%%%%%%%
SC gratefully acknowledges the Science Foundation Ireland Starting Investigator Research Grant ``SpeedDemon" (No. 18/SIRG/5508) for financial support.
%%%%%%%%%%%%%%%%%%%%%
NTZ acknowledges support from the Independent Research Fund Denmark, the Carlsberg Foundation, and the Jens Chr. Skou fellowship program funded by the Aarhus University Research Foundation. 
%%%%%%%%%%%%%%%%%%%%%
B.\c{C}. acknowledges support from The Research Fund of Bah\c{c}e\c{s}ehir University (BAUBAP).
%%%%%%%%%%%%%%%%%%%%%%
The authors would like to thank \"{O}zg\"{u}r E. M\"{u}stecapl{\i}o\u{g}lu for useful discussions.

%%%%%%%%%%%%%%%%%%%%%%%%%%%%%
\appendix
%%%%%%%%%%%%%%%%%%%%%%%%%%%%%

%%%%%%%%%%%%%%%%%%%%%%%%%%%%%

\section{Solution of the dynamics in Eq.~\eqref{LindEqSD}}\label{LindEqSDSol}
Consider the system of differential equations
\begin{equation}
\dot{\varrho}^{2}_{\text{sd}}(t) = -\Gamma_{21}\varrho^{2}_{\text{sd}}(t) + \Gamma_{32}\varrho^{3}_{\text{sd}}(t),~~
\dot{\varrho}^{3}_{\text{sd}}(t) = - \Gamma_{32}\varrho^{3}_{\text{sd}}(t) \text{ . }
\end{equation}
We can use the Laplace transform to solve the above equations. By denoting $\chi^{n}_{\text{sd}}(s)$ as the Laplace transformation of $\varrho^{n}_{\text{sd}}(t)$, we find the system of linear equations given by
\begin{align}
s\chi^{2}_{\text{sd}}(s) &= -\Gamma_{21}\chi^{2}_{\text{sd}}(s) + \Gamma_{32}\chi^{3}_{\text{sd}}(s) + \varrho^{n}_{\text{c}} \nonumber \\
s\chi^{3}_{\text{sd}}(s) &= - \Gamma_{32}\chi^{3}_{\text{sd}}(s) + \varrho^{n}_{\text{c}} \text{ , }
\end{align}
where we already used the initial conditions $\varrho^{n}_{\text{sd}}(0) = \varrho^{n}_{\text{c}}$, so that the solution for $\chi^{2}_{\text{sd}}(s)$ and $\chi^{3}_{\text{sd}}(s)$ are
\begin{equation}
\chi^{2}_{\text{sd}}(s) = \frac{(s+\Gamma_{32})\varrho^{2}_{\text{c}}+\Gamma_{32}\varrho^{3}_{\text{c}}}{(s+\Gamma_{32})(s+\Gamma_{21})},~~
\chi^{3}_{\text{sd}}(s) = \frac{\varrho^{3}_{\text{c}}}{s + \Gamma_{32}} \text{ . }
\end{equation}
Finally, we use the inverse transformation and get
\begin{align}
\varrho^{2}_{\text{sd}}(t) &= \frac{e^{-t\Gamma_{32}}\Gamma_{32}\varrho^{3}_{\text{c}} + e^{-t\Gamma_{21}}\left[ \Gamma_{21}\varrho^{2}_{\text{c}} - \Gamma_{32} \left(\varrho^{2}_{\text{c}}+\varrho^{3}_{\text{c}}\right)\right]}{\Gamma_{21}-\Gamma_{32}} \nonumber \\
\varrho^{3}_{\text{sd}}(t) &= e^{-t\Gamma_{32}}\varrho^{3}_{\text{c}} \text{ . }
\end{align}
Therefore, by using the case in which the battery is fully charged initially, where the initial conditions are $\varrho^{1}_{\text{c}}=\varrho^{2}_{\text{c}}=0$ and $\varrho^{3}_{\text{c}}=1$, one gets
\begin{equation}
\varrho^{2}_{\text{sd}}(t) = \frac{\Gamma_{32}e^{-t\Gamma_{32}} - \Gamma_{32} e^{-t\Gamma_{21}}}{\Gamma_{21}-\Gamma_{32}},~~
\varrho^{3}_{\text{sd}}(t) = e^{-t\Gamma_{32}} \text{ . }
\end{equation}

%\bibliography{/home/cs/Dropbox/School/Articles/Gaveta/Models/Bibliografia/mybib-noURL.bib}
\bibliography{Bib_QB}

\end{document}